\documentclass[12pt]{article}

\usepackage{amssymb}
\usepackage{amsmath}
\usepackage{amsfonts}

\newcommand{\R}{\mathbb{R}}
\newcommand{\C}{\mathbb{C}}
\newcommand{\Z}{\mathbb{Z}}

\newcommand{\be}{\begin{equation}}
\newcommand{\bea}{\begin{eqnarray}}
\newcommand{\eea}{\end{eqnarray}}
\newcommand{\nn}{\nonumber}
\newcommand{\kt}{\rangle}
\newcommand{\br}{\langle}

\newcommand{\ed}{\end{document}}
\newcommand{\bbr}{\br\!\br}
\newcommand{\kkt}{\kt\!\kt}

\newcommand{\pbr}{\prec}
\newcommand{\pkt}{\succ}

\begin{document}

\title{Exact PT-Symmetry Is Equivalent to Hermiticity.}
\author{Ali Mostafazadeh\thanks{E-mail address:
amostafazadeh@ku.edu.tr}\\ \\
Department of Mathematics, Ko\c{c} University,\\
Rumelifeneri Yolu, 34450 Sariyer,\\
Istanbul, Turkey}
\date{ }
\maketitle

\begin{abstract}
We show that a quantum system possessing an exact antilinear
symmetry, in particular $PT$-symmetry, is equivalent to a quantum
system having a Hermitian Hamiltonian. We construct the unitary
operator relating an arbitrary non-Hermitian Hamiltonian with
exact $PT$-symmetry to a Hermitian Hamiltonian. We apply our
general results to $PT$-symmetry in finite-dimensions and give the
explicit form of the above-mentioned unitary operator and
Hermitian Hamiltonian in two dimensions. Our findings lead to the
conjecture that non-Hermitian $CPT$-symmetric field theories are
equivalent to certain nonlocal Hermitian field theories.
\end{abstract}



The interest in $PT$-symmetric quantum mechanics \cite{bender} has
its origin in the idea that since the $CPT$ theorem follows from
the axioms of local quantum field theory, one might obtain a more
general field theory by replacing the axiom of the Hermiticity of
the Hamiltonian by the requirement of $CPT$-symmetry. The simplest
nonrelativistic example of such theories is the $PT$-symmetric
quantum mechanics. During the past five years there have appeared
dozens of publications exploring the properties of the
$PT$-symmetric Hamiltonians. Among these is a series of articles
\cite{p1}-\cite{p8} by the present author that attempt to
demonstrate that $PT$-symmetry can be understood most conveniently
using the theory of pseudo-Hermitian operators (See also
\cite{others}.) The recent articles of Bender, Meisimger, and Wang
\cite{bmw1,bmw2}, however, show that the mystery associated with
$PT$-symmetry has surprisingly survived the comprehensive
treatment offered by pseudo-Hermiticity. The aim of the present
article is to provide a conclusive proof that the exact
$PT$-symmetry is equivalent to Hermiticity. In particular, we
offer a complete treatment of $PT$-symmetry in finite-dimensions
that clarifies some of the issues raised in \cite{bmw2} and shows
that some of the claims made in \cite{bmw2} are not true. We also
comment on the nature and possible advantages of non-Hermitian
$CPT$-symmetric field theories.

First, we wish to point out that the results of \cite{bmw1}
regarding the $PT$-symmetry of Hermitian Hamiltonians follows from
the more general result that any diagonalizable pseudo-Hermitian
Hamiltonian is $PT$-symmetric where the $P$ and $T$ are the
generalized parity and time-reversal operators, \cite{p7}. The
definition of $P$ and $T$ used in \cite{bmw1} are originally given
for arbitrary diagonalizable pseudo-Hermitian Hamiltonians in
\cite{p7}; they are Eqs.~(77) and (78) of \cite{p7}. The statement
that any Hermitian Hamiltonian is $PT$-symmetric is actually not
surprising at all. A simple corollary of Theorem~2 of \cite{p3} is
that any Hermitian Hamiltonian has an antilinear symmetry. The
proof of this theorem provides an explicit construction of such
symmetries. Among them are the (generalized) $PT$ and $CPT$
symmetries that are considered in great detail in \cite{p7}.

We start our analysis by considering a linear operator $H'$ that
acts in a complex vector space $V$ and commutes with an invertible
antilinear operator ${\cal X}'$. Then as shown in \cite{p2}, the
eigenvalues of $H'$ are either real or come in complex-conjugate
pairs. Furthermore, if we demand that all the eigenvectors of $H'$
are also eigenvectors of ${\cal X}'$, i.e., the symmetry generated
by ${\cal X}'$ is exact, then the eigenvalues of $H'$ are
necessarily real. Now, let ${\cal H}$ be the (invariant) subspace
of $V$ spanned by the eigenvectors of $H'$. Then by construction
the restriction $H$ of $H'$ to ${\cal H}$ will be diagonalizable,
and the restriction ${\cal X}$ of ${\cal X}'$ to ${\cal H}$ will
generate an exact symmetry of $H$.

Next, suppose that $\br~~,~~\kt$ is an arbitrary complete
positive-definite inner product on ${\cal H}$, so that ${\cal H}$
is endowed with the structure of a separable Hilbert space. Then
$H$ is a diagonalizable operator acting in ${\cal H}$ and having a
real spectrum. We will identify it as the Hamiltonian of a
physical system whose state vectors belong to ${\cal H}$. The
dynamics of the system is then determined by the Schr\"odinger
equation
    \be
    i\hbar\frac{d}{dt}\psi(t)=H\psi(t).
    \label{sch-eq}
    \end{equation}
It is tempting to view ${\cal H}$ as the Hilbert space for this
quantum system. However, in general, $H$ is not a Hermitian
operator with respect to the inner product $\br~~,~~\kt$ of ${\cal
H}$. Hence, the time-evolution generated by $H$ in ${\cal H}$ will
not be unitary.

If we assume that $H$ has a discrete spectrum, then according to
Theorem~3 of \cite{p3} $H$ is Hermitian with respect to a
positive-definite inner product $\bbr~~,~~\kkt$ on ${\cal H}$.
Like any other inner product on ${\cal H}$, $\bbr~~,~~\kkt$ will
have the form \cite{kato}:
    \be
    \bbr\psi,\phi\kkt=\br\psi,\eta_+\phi\kt,~~~~~~~~~~\forall
    \psi,\phi\in{\cal H},
    \label{inner1}
    \end{equation}
for some Hermitian, invertible, linear operator $\eta_+:{\cal
H}\to{\cal H}$. The Hermiticity of $H$ with respect to
$\bbr~~,~~\kkt$, i.e.,
    \be
    \bbr\psi,H\phi\kkt=\bbr H\psi,\phi\kkt,~~~~~~~~~~
    \forall\psi,\phi\in{\cal H},
    \label{herm}
    \end{equation}
is equivalent to its $\eta_+$-pseudo-Hermiticity \cite{p1}:
    \be
    H^\dagger=\eta_+H\eta_+^{-1}.
    \label{ph}
    \end{equation}
Moreover, the fact that $\bbr~~,~~\kkt$ is a positive-definite
inner product implies that $\eta_+$ is a positive-definite
operator. This in turn means that $\eta_+$ has a positive-definite
square root $\rho_+$, i.e., there exists a positive Hermitian
operator $\rho_+:{\cal H}\to{\cal H}$ such that
    \be
    \eta_+=\rho_+^2.
    \label{eta=rho}
    \end{equation}
Clearly, $\rho_+$ is invertible.

Next, let $\tilde{\cal H}$ denote the span of the eigenvectors of
$H$ endowed with the inner product $\bbr~~,~~\kkt$. As a vector
space $\tilde{\cal H}$ coincides with ${\cal H}$. Therefore we may
view $\rho_+$ as a linear invertible operator mapping $\tilde{\cal
H}$ onto ${\cal H}$. We can easily show that for all
$\psi,\phi\in{\cal H}$,
    \[\bbr\rho_+^{-1}\psi,\rho_+^{-1}\phi\kkt=
    \br\rho_+^{-1}\psi,\eta_+\rho_+^{-1}\phi\kt=
    \br\psi,\rho_+^{-1}\eta_+\rho_+^{-1}\phi\kt=
    \br\psi,\phi\kt.\]
Equivalently, we have for all $\phi\in{\cal H}$ and
$\tilde\psi\in\tilde{\cal H}$,
    \[ \bbr\tilde\psi,\rho_+^{-1}\phi\kkt=\br\rho_+
    \tilde\psi,\phi\kt.\]
Comparing this equation with the defining relation for
$\rho_+^{-1\dagger}$, namely
$\bbr\tilde\psi,\rho_+^{-1}\phi\kkt=\br\rho_+^{-1\dagger}
\tilde\psi,\phi\kt$, we see that
$\rho_+^{-1\dagger}=\rho_+=(\rho_+^{-1})^{-1}$. Therefore,
$\rho_+^{-1}:{\cal H}\to\tilde{\cal H}$ is a unitary operator; the
Hilbert spaces ${\cal H}$ and $\tilde{\cal H}$ are related by a
unitary operator. In particular, for every Hamiltonian operator
$h$ defining a time-evolution in ${\cal H}$, we may define a
Hamiltonian
    \be
    \tilde h:=\rho_+^{-1}h\rho_+,
    \label{can-0}
    \end{equation}
acting in $\tilde{\cal H}$ such that under the action of
$\rho_+^{-1}$ the solutions $\psi(t)$ of the Schr\"odinger
equation for the Hamiltonian $h$ are mapped to the solutions
$\tilde\psi(t)$ of the Schr\"odinger equation for $\tilde h$. The
observables $O:{\cal H}\to{\cal H}$ are also mapped to the
observables $\tilde O:\tilde{\cal H}\to\tilde{\cal H}$ by the
unitary similarity transformation:
    \be
    \tilde O= \rho_+^{-1}O\rho_+.
    \label{O}
    \end{equation}

Now, if we set $\tilde h=H$, i.e., view $H$ as a Hamiltonian
acting in the Hilbert space $\tilde{\cal H}$, then
    \be
    h:=\rho_+H\rho_+^{-1},
    \label{can}
    \end{equation}
will be a Hermitian Hamiltonian acting in the original Hilbert
space ${\cal H}$. The Hermiticity of $h$ follows from the fact
that $H$ is Hermitian with respect to the inner product
$\bbr~~,~~\kkt$ on $\tilde{\cal H}$, and that $\rho_+^{-1}$ is
unitary.

By construction, the Hamiltonians $H$ and $h$ are related by a
unitary transformation mapping two different Hilbert spaces with
the same vector space structure. Using the terminology of
\cite{p3}, we say that the quantum systems determined by $({\cal
H},h)$ and $(\tilde{\cal H},H)$ are related by a pseudo-canonical
transformation. Clearly, they are physically equivalent.

In summary, we have shown that if a quantum system has an exact
antilinear symmetry (with an invertible symmetry generator) then
one can describe the same system using a Hermitian Hamiltonian.
This applies to $PT$-symmetric systems whose generator is clearly
invertible.

The construction of the unitary operator $\rho_+^{-1}$ requires
the knowledge of the eigenvectors of the Hamiltonian $H$. If the
Hilbert space is an infinite-dimensional function space and $H$ is
a differential operator, then $\rho_+^{-1}$ and consequently the
Hamiltonian $h$ are in general nonlocal (non-differential)
operators. This suggests that the idea of replacing the
Hermiticity condition on the Hamiltonian of a local quantum field
theory by its $CPT$-symmetry will probably give rise to a theory
which is equivalent to a nonlocal field theory with a Hermitian
Hamiltonian. This should not however overshadow the importance of
this idea as it suggests the possibility of treating certain
nonlocal field theories using equivalent local CPT-symmetric field
theories with non-Hermitian Hamiltonians.

In the following we explore the utility of our findings in the
study of $PT$-symmetry in finite dimensions \cite{bmw2}, where
${\cal H}=\C^D$ for some $D\in\Z^+$.

In \cite{bmw2}, the authors explore certain matrix Hamiltonians
that they identify with the finite-dimensional analogs of the
Hamiltonians studied within the context of $PT$-symmetric quantum
mechanics \cite{bender}. The analysis of \cite{bmw2} involves
considering complex symmetric Hamiltonians $H$ that admit an
antilinear symmetry generated by ${\cal X}:=PT$ where $P$ is a
real symmetric matrix satisfying $P^2=1$, i.e., it is an
involution, and $T$ is complex-conjugation $\star$ (for all
$\psi\in{\cal H}$, $\star\psi:=\psi^*$.) They outline a
construction of the most general real symmetric matrix $P$ which
is an involution, impose the condition that $H$ commutes with
$PT$, restrict to the range of parameters of $H$ where the
$PT$-symmetry is exact, and define the indefinite $PT$-inner
product,
    \be
    (\psi|\phi):=[PT\psi]^T\cdot\phi,
    \label{PT}
    \end{equation}
where $~^T$ stands for the transpose and a dot means matrix
multiplication. For the case $D=2$, they compute the eigenvectors
of $H$, introduce a charge-conjugation operator $C$, such that $H$
commutes with $C$ and consequently $CPT$, and show that the
$CPT$-inner product,
    \be
    \br\psi|\phi\kt:=[CPT\psi_a]^T\cdot\phi,
    \label{CPT}
    \end{equation}
is positive-definite. Among the statements made in \cite{bmw2} are
    \begin{itemize}
    \item[]
    Claim 1: A finite-dimensional $PT$-symmetric Hamiltonian
    (which is a certain complex symmetric matrix) is not unitarily
    equivalent to any Hermitian matrix Hamiltonian.
   \item[]
    Claim 2: The extension to non-symmetric $PT$-symmetric matrix
    Hamiltonians cannot be pursued by the methods of the theory of
    pseudo-Hermitian operators as outlined in \cite{p7}, because
    they lead to nonunitary evolutions.
    \end{itemize}
In the remainder of this article we show how the general results
described above explain the findings reported in \cite{bmw2},
prove that the claims 1.\ and 2.\ are false, and discuss an
extension of the results of \cite{bmw2} on $PT$-symmetry in
finite-dimensions to nonsymmetric matrix Hamiltonians.

First, we use the fact that $T$ is complex-conjugation and $P$ is
a real symmetric involution to show that $(PT)^2=1$. This together
with the observation that $H$ is a $PT$-symmetric symmetric
complex matrix imply
    \[H=PT~H~PT=P(T H) TP=P~H^*P=PH^\dagger P.\]
Multiplying both sides of this equation from left and right by $P$
and using $P^2=1$, we have
    \be
    H^\dagger=P~H~P=P~H~P^{-1}.
    \label{ph-2}
    \end{equation}
Hence $H$ is $P$-pseudo-Hermitian.

Next, let $\pbr~~,~~\pkt$ denote the ordinary Euclidean inner
product on ${\cal H}=\C^D$, i.e.,
    \be
    \pbr\psi,\phi\pkt:=\psi^\dagger\cdot\phi,~~~~~~~~~~~\forall
    \psi,\phi\in{\cal H},
    \label{euclid}
    \end{equation}
where $\psi^\dagger:=\psi^{T*}$. Then in view of Eqs.~(\ref{PT})
and (\ref{euclid}), and the fact that $P$ is real and symmetric,
    \[(\psi|\phi)=[P\psi^*]^T\cdot\phi=
    \psi^{T*}P\cdot\phi=\pbr\psi,P\phi\pkt,~~~~~~~~~~~\forall
    \psi,\phi\in{\cal H}.\]
Therefore the $PT$-inner product of \cite{bmw2} is just the
pseudo-inner product \cite{p1}:
    \[\bbr\psi,\phi\kkt_{\eta}:=\pbr\psi,\eta\phi\pkt,~~~~~~~~~~~~
    \forall\psi,\phi\in{\cal H},\]
corresponding to the choice $\eta=P$.

Because the eigenvalues of $H$ are real, it is
$\eta_+$-pseudo-Hermitian for a positive-definite operator
$\eta_+$, i.e., (\ref{ph}) holds. As discussed in \cite{p1,p7}, if
we introduce $C:=\eta_+^{-1}P$, we can use Eqs.~(\ref{ph}) and
(\ref{ph-2}) to show
    \be
    [H,C]=H\eta_+^{-1}P-\eta_+^{-1}PH=\eta_+^{-1}H^\dagger P-
    \eta_+^{-1}H^\dagger P=0,
    \label{C}
    \end{equation}
i.e., $C$ is a linear symmetry generator. Furthermore, if we
repeat the arguments leading to Eq.~(75) of \cite{p7} we find that
the $CPT$-inner product is nothing but the $\eta_+$-inner product:
    \[\br\psi|\phi\kt=\pbr\psi,\eta_+\psi\pkt=:
    \bbr\psi,\phi\kkt.\]

As a concrete example, we give an explicit construction of
$\eta_+$, $P$, and $C$ for the $2\times 2$ Hamiltonians studied in
\cite{bmw2}, namely
    \be
    H=\left(\begin{array}{cc}
    r+t\cos\varphi-is\sin\varphi & is\cos\varphi+t\sin \varphi\\
    is\cos\varphi+t\sin \varphi&r-t\cos\varphi+is\sin\varphi
    \end{array}\right),
    \label{H}
    \end{equation}
where $r,s,t,\varphi$ are real parameters and
    \be
    |s|\leq|t|.
    \label{condi-1}
    \end{equation}
We will also compute the Hermitian matrix $h$ that is unitarily
equivalent to $H$.

As pointed out in \cite{bmw2},
    \be
    \psi_n:=\left(\begin{array}{c}
    a_n \cos\,\frac{\varphi}{2}+ib_n\sin\,\frac{\varphi}{2}\\
    a_n\sin\,\frac{\varphi}{2}-ib_n\cos\,\frac{\varphi}{2}
    \end{array}\right),~~~~~~~~~~~~~~~n=\pm,
    \label{eg-ve}
    \end{equation}
with
    \be
    a_n:=\frac{\sin\alpha}{\sqrt{2(1-n\cos\alpha)\cos\alpha}},~~~~~~~~~~~
    b_n:=\frac{(-1+n\cos\alpha)}{\sqrt{2(1-n\cos\alpha)\cos\alpha}},
    \label{a-b}
    \end{equation}
and $\alpha:=\sin^{-1}(s/t)\in(-\pi/2,\pi/2)$, are linearly
independent eigenvectors of $H$.

We also notice that because $H$ is symmetric, $H^\dagger=H^*$.
Hence $\psi_n^*$ are eigenvectors of $H^\dagger$. If we let
$\phi_n=n\psi_n^*$, we find a pair of linearly independent
eigenvectors of $H^\dagger$, namely
    \be
    \phi_n=n\left(\begin{array}{c}
    a_n \cos\,\frac{\varphi}{2}-ib_n\sin\,\frac{\varphi}{2}\\
    a_n\sin\,\frac{\varphi}{2}+ib_n\cos\,\frac{\varphi}{2}\end{array}\right),
    \label{eg-ve2}
    \end{equation}
that together with $ \psi_n$ form a complete biorthonormal system
$\{\psi_n,\phi_n\}$ for the Hilbert space ${\cal H}=\C^2$. That is
they satisfy
    \be
    \pbr\phi_n,\psi_m\pkt=\phi_n^\dagger\cdot\psi_m=\delta_{mn}
    ,~~~~~~~
    \psi_+\cdot\phi_+^\dagger+\psi_-\cdot\phi_-^\dagger=I,
    \label{biorth}
    \end{equation}
where $I$ is the identity matrix.

Now, we can compute the positive-definite operator $\eta_+$, and
the generalized parity ${\cal P}$ and charge conjugation ${\cal
C}$ operators as defined in \cite{p7}, namely
    \bea
    \eta_+&:=&\phi_+\cdot\phi_+^\dagger+\phi_-\cdot\phi_-^\dagger,
    \label{e1}\\
    {\cal P}&:=&\phi_+\cdot\phi_+^\dagger-\phi_-\cdot\phi_-^\dagger,
    \label{e2}\\
    {\cal
    C}&:=&\psi_+\cdot\phi_+^\dagger-\psi_-\cdot\phi_-^\dagger.
    \label{e3}
    \eea
Substituting (\ref{eg-ve}) and (\ref{eg-ve2}) in these equations,
we find
    \bea
    \eta_+&=&
    \left(\begin{array}{cc}
    {\rm sec}\:\alpha & i\tan\alpha\\
    -i\tan\alpha&{\rm sec}\:\alpha\end{array}\right),
     \label{e4}\\
       {\cal P}&=&\left(\begin{array}{cc}
    \cos\varphi & \sin\varphi\\
    \sin\varphi&-\cos\varphi\end{array}\right),
     \label{e5}\\
       {\cal C}&=&\left(\begin{array}{cc}
    {\rm sec}\:\alpha\;\cos\varphi-i\tan\alpha\;\sin\varphi &
    {\rm sec}\:\alpha\;\sin\varphi+i\tan\alpha\;\cos\varphi \\
    {\rm sec}\:\alpha\;\sin\varphi+i\tan\alpha\;\cos\varphi &
    -{\rm sec}\:\alpha\;\cos\varphi+i\tan\alpha\;\sin\varphi
    \end{array}\right).
    \label{e6}
    \eea
One can directly check that indeed $\eta_+$ and $H$ satisfy
(\ref{ph}), i.e., $H$ is $\eta_+$-pseudo-Hermitian, and that the
eigenvalues of $\eta_+$, which are given by ${\rm
sec}~\alpha\pm\tan\alpha=\sqrt{1+\tan^2\alpha}\pm\tan\alpha$, are
positive. Moreover, Eqs.~(\ref{e5}) and (\ref{e6}) are identical
with the expressions for the parity $P$ and the charge-conjugation
$C$ given in \cite{bmw2}. We have obtained them by a systematic
application of the general results of \cite{p7}.

Next, we show that contrary to the claims of \cite{bmw2} the
quantum system defined by Hamiltonian (\ref{H}) is equivalent to a
quantum system having a Hermitian Hamiltonian $h$. For this
purpose we calculate the positive square root $\rho_+$ of
$\eta_+$. The result is
    \be
    \rho_+=\left(\begin{array}{cc}
        r_+&-i r_-\\
    i r_-&r_+
    \end{array}\right),
    \label{rho=}
        \end{equation}
where
    \[ r_\pm:=\frac{1}{2}\,\left(\sqrt{ {\rm
    sec}~\alpha-\tan\alpha}\pm
    \sqrt{ {\rm sec}~\alpha+\tan\alpha}\right).\]
Inserting (\ref{H}) and (\ref{rho=}) in (\ref{can}), we obtain
    \be
    h=\left(\begin{array}{cc}
    r+\sqrt{t^2+s^2}\,\cos\varphi&\sqrt{t^2+s^2}\,\sin\varphi\\
    \sqrt{t^2+s^2}\,\sin\varphi&r-\sqrt{t^2+s^2}\,\cos\varphi
    \end{array}\right)=r I+\sqrt{t^2+s^2}\;{\cal P}.
    \label{tilde-H}
    \end{equation}
This Hamiltonian is a real symmetric matrix, so it is Hermitian as
expected. We also see that it is ${\cal P}$-symmetric.

A quantum system described by the Hamiltonian $H$ that is viewed
as acting in the Hilbert space ${\cal H}$ obtained by endowing
$\C^2$ with the inner product (\ref{inner1}), which is the same as
the $CPT$-inner product, may be equally well described by the
Hermitian Hamiltonian $h$ viewed as acting in $\C^2$ endowed with
the Euclidean inner product (\ref{euclid}). There is simply no
advantage in considering the Hamiltonians of the form (\ref{H})
and imposing the condition that they should generate a unitary
time-evolution. This condition leads one to the study of the
well-understood two-level Hermitian Hamiltonians \cite{nova}.

Next, we wish to point out that the most general $PT$-symmetric
matrix Hamiltonians (with $PT$ to be understood as the generalized
parity-time-reversal operator \cite{p7}) are the pseudo-Hermitian
matrices. Among these are the quasi-Hermitian Hamiltonians
\cite{quasi,p7} that have an unbroken $PT$-symmetry. But these are
related to Hermitian Hamiltonians via similarity transformations
by invertible matrices. Each matrix Hamiltonian (acting in $\C^D$
and) having an exact $PT$-symmetry lives in an orbit of the
adjoint action of the group $GL(D,\C)$ on the $u(D)$ subalgebra of
the Lie algebra ${\cal G}\ell(D,\C)$. In particular it is
diagonalizable. More general $PT$-symmetric Hamiltonians may or
may not be diagonalizable. Because the exponential of $i$ times a
pseudo-Hermitian matrix is necessarily pseudo-unitary and all the
pseudo-unitary matrices are obtained in this way \cite{p8}, one
can use the general characterization of pseudo-unitary matrices
given in \cite{p8} to determine the number of the independent real
parameters in the most general pseudo-Hermitian matrix. Note
however that if a pseudo-Hermitian matrix has a broken
$PT$-symmetry so that it has complex eigenvalues or it is not
diagonalizable, then it cannot be used as a Hamiltonian capable of
supporting a unitary evolution. In this case one can easily show
that there is no positive-definite inner product in which this
Hamiltonian is Hermitian. This in turn means \cite{p1} that for
any choice of positive-definite inner product on $\C^D$, there are
solutions of the Schr\"odinger equation whose norm will depend on
time.

For $D\times D$ matrix Hamiltonians with exact $PT$-symmetry one
can easily count the maximum number of free real parameters. But
what is important is the number of independent parameters
corresponding to physically distinct Hamiltonians. The similarity
transformations by invertible matrices are essentially gauge
transformations relating physically equivalent Hamiltonians.
Therefore, there are as many physically distinct $D\times D$
matrix Hamiltonians with exact $PT$-symmetry as physically
distinct Hermitian $D\times D$ matrix Hamiltonians. The latter
have at most $D^2$ real parameters.\footnote{One can diagonalize a
Hermitian Hamiltonian by a unitary transformation and transform it
into a traceless matrix by a time-dependent phase transformation
of the state vectors, \cite{nova}. This implies that distinct
unitary physical systems having $D\times D$ matrix Hamiltonian are
uniquely determined by $D-1$ free parameters. These may be
identified with the transition energies.} The fact that the
authors of \cite{bmw2} obtain a smaller number is because they
confine their study to symmetric complex matrices. As we argued
above one can consistently apply the results of \cite{p7} to
consider nonsymmetric $PT$-symmetric Hamiltonians that support
unitary evolutions provided that the $PT$-symmetry is not broken.

In \cite{p6}, we provide a complete analysis of general $2\times
2$ pseudo-Hermitian Hamiltonians. In particular, we show that the
number of free parameters in a traceless diagonalizable $2\times
2$ pseudo-Hermitian Hamiltonian having real eigenvalues is 5.
Allowing for a nonzero trace is equivalent to adding a
pseudo-Hermitian matrix that is proportional to the identity
matrix, i.e., $a_0 I$ for some $a_0\in\R$. Hence the number of
free real parameters in the most general diagonalizable $2\times
2$ pseudo-Hermitian Hamiltonian $H$ with real eigenvalues is 6. As
we show in \cite{p7}, we can construct the generalized parity
${\cal P}$, time-reversal ${\cal T}$, and charge-conjugation
${\cal C}$ operators and show that $H$ has ${\cal PT}$-, ${\cal
C}$-, and ${\cal CPT}$-symmetries. These symmetry generators are
involutions, i.e., $({\cal PT})^2={\cal C}^2=({\cal CPT})^2=I$.
However, the operators ${\cal P}$ and ${\cal T}$ are involutions
(${\cal P}^2={\cal T}^2=I$) provided that the eigenvectors of $H$
and $H^\dagger$ fulfill certain conditions (See statement~6 of
Lemma~1 in \cite{p7}.) In the following, instead of trying to
satisfy these conditions, we will first identify $T$ with
complex-conjugation $\star$ and use a direct method to construct
the most general $2\times 2$ matrix Hamiltonian admitting an exact
$PT$-symmetry for an indefinite Hermitian involution $P$ (This
means that $P^\dagger=P=P^{-1}$ and $P$ has real eigenvalues of
opposite sign.),  so that $PT$ is also an involution. We will then
extend our analysis to the most general case where $T$ is an
arbitrary Hermitian, antilinear, involution.

Let $T=\star$, then the equation $(PT)^2=I$ may be written as
$P=\star P\star=P^*$. Therefore, $P$ is a real
Hermitian (equivalently a real symmetric) matrix. Moreover, the
condition that $P$ is an indefinite involution implies that its eigenvalues
are $\pm 1$. This is sufficient to establish that
    \be
    P=O\sigma_3 O^{-1},
    \label{e18}
    \end{equation}
where $\sigma_3$ is the diagonal Pauli matrix (See (\ref{pauli})
below.) and $O$ is some special orthogonal matrix, i.e., $O\in
SO(2)$. $O$ is in particular unitary and as any other unitary
matrix may be written as the exponential of $i$ times a linear
combination of the Pauli matrices:
    \be
    \sigma_1=\left(\begin{array}{cc}
    0 & 1\\
    1 & 1\end{array}\right),~~~~~~
    \sigma_2=\left(\begin{array}{cc}
    0 & -i\\
    i & 0\end{array}\right),~~~~~~
     \sigma_3=\left(\begin{array}{cc}
    1 & 0\\
    0 & -1\end{array}\right).
    \label{pauli}
    \end{equation}
The fact that $O$ is a real matrix then implies that it has the
form
    \be
    O=e^{-i\varphi\sigma_2/2}
    \label{e19}
    \end{equation}
for some $\varphi\in[0,2\pi)$. Inserting (\ref{e19}) in (\ref{e18}),
we have
    \be
    P=e^{-i\varphi\sigma_2/2}\sigma_3\; e^{i\varphi\sigma_2/2}=
    \cos\varphi\;\sigma_3+\sin\varphi\;\sigma_1.
    \label{e30}
    \end{equation}
To establish the second equality in (\ref{e30}) we used the
identity \cite{nova}:
    \be
    e^{-i\vartheta\sigma_i/2}\;\sigma_j\;
    e^{i\vartheta\sigma_2/2}=\cos\vartheta\;\sigma_j+\sin\vartheta\;
    \sum_{k=1}^3\epsilon_{ijk}\sigma_k,~~~~~~~~~\forall i\neq j,
    \label{identity}
    \end{equation}
where $\vartheta\in\C$ and $\epsilon_{ijk}$ is the totally
antisymmetric Levi-Civita symbol with $\epsilon_{123}=1$.
Eq.~(\ref{e30}) is identical to (\ref{e5}), as expected
\cite{bmw2}.

Next, we note that the $PT$-symmetry of $H$ (i.e., $[H,PT]=0$)
together with $(PT)^2=1$ and $P^2=1$ imply
    \be
    H^*=PHP.
    \label{e20}
    \end{equation}
Defining
    \be
    H_0:=O^{-1} H O,
    \label{e21}
    \end{equation}
and using (\ref{e18}), (\ref{e19}), and (\ref{e20}), we find
    \be
    H_0^*=\sigma_3\; H_0\, \sigma_3.
    \label{h-star}
    \end{equation}
As any $2\times 2$ complex matrix, $H_0$ may be written as a
linear combination of the identity matrix and the Pauli matrices,
$H_0=a_0 I+\sum_{i=1}^3 a_i\sigma_i$, with $a_0,a_1,a_2,a_3\in\C$.
Substituting this equation in (\ref{h-star}) and using
(\ref{pauli}) we see that $a_0,a_2,a_3$ must be real and $a_1$
must be imaginary. Letting $\alpha_1:=i a_1\in\R$, we then have
    \be
    H_0=a_0I-i\alpha_1\sigma_1+a_2\sigma_2+a_3\sigma_3=
    \left(\begin{array}{cc}
    a_0+a_3 & -i(\alpha_1+a_2)\\
    i(-\alpha_1+a_2) & a_0-a_3\end{array}\right).
    \label{h=2}
    \end{equation}
Next, we use (\ref{e19}), (\ref{identity}), (\ref{e21}) and
(\ref{h=2}) to compute
    \be
    H=O\,H_0\,O^{-1}=a_0I-(a_3\sin\varphi+i\alpha_1\cos\varphi)
    \sigma_1+a_2\sigma_2+(a_3\cos\varphi-i\alpha_1\sin\varphi)
    \sigma_3
    \label{H=2}
    \end{equation}
If we relabel the parameters according to
    \[ r:=a_0,~~~~~~s:=-\alpha_1,~~~~~~~t:=a_3,~~~~~~u:=a_2,\]
and insert (\ref{pauli}) in (\ref{H=2}), we obtain
    \be
    H= \left(\begin{array}{cc}
    r+t\cos\varphi-is\sin\varphi &
    t\sin\varphi+i(s\cos\varphi-u)\\
    t\sin\varphi+i(s\cos\varphi+u)&
    r-t\cos\varphi+si\sin\varphi\end{array}\right).
    \label{H=3}
    \end{equation}
As seen from this equation, $H$ has 5 free real parameters. The
condition of the exactness of the $PT$-symmetry implies that the
eigenvalues of $H$ are real. These are also the eigenvalues of
$H_0$. The fact that the eigenvalues of $H_0$ are real implies
that the determinant of the traceless part of $H_0$ must be either
zero or negative. In view of (\ref{h=2}) this yields $s^2-t^2-u^2=
\alpha_1^2-a_2^2-a_3^2\leq 0$. If $s^2-t^2-u^2=0$, either
$H=H_0=0$ or $H_0$ and consequently $H$ are not diagonalizable
\cite{p6}. Requiring that $H$ is a nonzero diagonalizable
Hamiltonian so that it supports a nontrivial (nonstationary)
unitary time-evolution (with respect to some positive-definite
inner product on $\C^2$) is equivalent to the condition
$s^2-t^2-u^2<0$, alternatively
    \be
    |s|<\sqrt{t^2+u^2}.
    \label{condi2}
    \end{equation}

Now, suppose that the condition~(\ref{condi2}) is satisfied so
that $H$ is diagonalizable and has real eigenvalues $E_\pm$. Let
$\psi_\pm$ be a pair of linearly independent eigenvectors of $H$,
so that $H\psi_\pm=E_\pm\psi_\pm$. Acting both sides of $[PT,H]=0$
on $\psi_\pm$, we can easily show that $PT\psi_\pm$ are also
eigenvectors of $H$ with eigenvalue $E_\pm$. There are two
possibilities:
    \begin{enumerate}
    \item $E_+=E_1$: Then $H$ is a real multiple of the identity
    matrix, i.e., $H=E_+ I$, and we can select $\psi_+=\pi_+$ and
    $\psi_-=i\pi_-$, where $\pi_\pm$ are a pair of eigenvectors of
    $P$ with eigenvalue $\pm 1$. It is obvious that $H\psi_\pm=
    E\psi_\pm$ and $PT\psi_\pm=P[\pm\psi_\pm]=\psi_\pm$.
    \item $E_+\neq E_-$: Then $PT\psi_\pm=N_\pm\psi_\pm$
    for some $N_\pm\in\C-\{0\}$, and we can always rescale the
    eigenvectors $\psi_\pm$ and choose the phases of $N_\pm$
    so that $PT\psi_\pm=\psi_\pm$.
    \end{enumerate}
This shows that (\ref{condi2}) is the necessary and sufficient
condition for the exactness of $PT$-symmetry.

The symmetric Hamiltonian (\ref{H}) and the condition
(\ref{condi-1}) ensuring its exact $PT$-symmetry are special cases
of the Hamiltonian~(\ref{H=3}) and the condition (\ref{condi2}).
Because the Hamiltonians (\ref{H}) and (\ref{H=3}) differ by
$u\;\sigma_2$, one may wonder if they are related by a unitary
similarity transformation. In order to see that this is indeed the
case, we let $t'\in\R^+$ and $\beta\in[0,2\pi)$ be given by
    \[ t':=\sqrt{t^2+u^2},~~~~~~~
    \sin\beta:=\frac{u}{\sqrt{t^2+u^2}},~~~~~~~
    \cos\beta:=\frac{t}{\sqrt{t^2+u^2}},\]
and introduce
    \bea
    H'&:=&\left(\begin{array}{cc}
    r+t'\cos\varphi-is\sin\varphi & is\cos\varphi+t'\sin \varphi\\
    is\cos\varphi+t'\sin \varphi&r-t'\cos\varphi+is\sin\varphi
    \end{array}\right),
    \label{H-prime}\\
    &&\nn\\
    U_1&:=&e^{i\varphi\sigma_2/2}\:e^{i\beta\sigma_1/2}\:
    e^{-i\varphi\sigma_2/2}.
    \label{U1}
    \eea
Then using (\ref{identity}) we can show that
    \bea
    H'&=&e^{i\varphi\sigma_2/2}
    (r\;I+is\;\sigma_1+t'\:\sigma_3)\;e^{-i\varphi\sigma_2/2},
    \label{H-prime2}\\
    H&=& U_1 H' U_1^{-1},
    \label{H=H-prime}
    \eea
where $H$ is the Hamiltonian (\ref{H=3}). Eq.~(\ref{H=H-prime})
indicates that the Hamiltonian (\ref{H=3}) may be mapped to a
symmetric Hamiltonian of the form (\ref{H}) by a unitary
transformation. A direct consequence of this observation is that
the Hamiltonians (\ref{H=3}) are also equivalent to the Hermitian
Hamiltonians (\ref{tilde-H}); in view of (\ref{can}) and
(\ref{H=H-prime}) we have
    \be
    H=U_2 h' U_2^{-1},
    \label{H=UH-tilde-U}
    \end{equation}
where $U_2:=U_1{\rho'_+}^{-1}$, $\rho_+'$ is the matrix
(\ref{rho=}) with $\alpha\in(-\pi/2,\pi/2)$ given by
$\alpha=\sin^{-1}(s/t')$, and $h'$ is the Hamiltonian
(\ref{tilde-H}) with $t$ replaced by $t'$. Because $U_1$ is a
unitary matrix, $U_2$ viewed as an operator mapping $\C^2$ endowed
with the inner product (\ref{inner1}) to $\C^2$ endowed with the
Euclidean inner product (\ref{euclid}) is unitary. Therefore,
Eq.~(\ref{H=UH-tilde-U}) establishes the unitary-equivalence of
the Hamiltonians~(\ref{H=3}) to the Hermitian Hamiltonians of the
form~(\ref{tilde-H}).

Next, we wish to construct the most general $2\times 2$
Hamiltonians admitting an exact $PT$-symmetry such that $P$ and
$T$ are general Hermitian (respectively linear and antilinear)
commuting involutions. To do this, we first recall that the
Hermiticity condition for an antilinear operator $T$ has the form
\cite{weinberg}
    \be
    \pbr\psi,T\phi\pkt=\pbr\phi,T\psi\pkt,
    ~~~~~~~~\forall\psi,\phi\in{\cal H}=\C^2,
    \label{herm2}
    \end{equation}
and that any antilinear operator acting in $\C^2$ may be expressed
as
    \be
    T=\tau\star
    \label{e40}
    \end{equation}
for some linear operator $\tau:\C^2\to\C^2$. Then imposing the
condition that $T$ is an involution, i.e., $T^2=1$ and using
(\ref{herm2}), we can show that
    \be
    \tau^\dagger=\tau^{-1}=\tau^*.
    \label{tau-tau}
    \end{equation}
In other words, $\tau$ is a complex, symmetric, unitary matrix.
Writing $\tau$ as the exponential of $i$ times a linear
combination of $I$ and the Pauli matrices and requiring that it is
symmetric yields the general form of $\tau$, namely
    \be
    \tau=e^{i\gamma}[\cos\xi\,I+
    i\sin\xi(\cos\zeta\,\sigma_1+\sin\zeta\,\sigma_3)],
    \label{tau=}
    \end{equation}
where $\gamma,\xi,\zeta\in[0,2\pi)$.

Next, we introduce the unitary symmetric matrix
    \be
    U:=e^{i\gamma/2}\;e^{i\xi(\cos\zeta\;\sigma_1+\sin\zeta\;\sigma_3)/2}.
    \label{U=5}
    \end{equation}
Then in view of the identity
    \[ e^{i\varrho\,\sum_{i=1}^3n_i\sigma_i}=\cos\varrho\;I+i\sin\varrho\;
    \sum_{i=1}^3 n_i\sigma_i,\]
where $\varrho\in\R$ and $n_i$ are the components of a unit vector
$\hat n\in\R^3$, we can check that
    \be
    \tau=U^2.
    \label{t=u}
    \end{equation}
Substituting this equation in (\ref{e40}) and making use of the fact that $U$
is both unitary and symmetric, so that $U^*=U^\dagger=U^{-1}$, we have
    \be
    T=U^2\star=U\;\star\;(\star\; U\;\star)=U\;\star\;U^*=U\;\star\;U^{-1}.
    \label{e60}
    \end{equation}

Eqs.~(\ref{e60}) reduce the analysis of the general $PT$-symmetric
$2\times 2$ Hamiltonians $H$ with $T$ given by (\ref{e40}) to that
of the Hamiltonians~(\ref{H=3}). In order to see this, we introduce
    \be
    \check T:=U^{-1} T\; U=\star,~~~~~~~~~~~~
    \check P:=U^{-1}P\;U,~~~~~~~~~~~~
    \check H:=U^{-1}H\;U.
        \label{e61}
    \end{equation}
In view of the fact that $U$ is a unitary matrix, it is easy to
see that $\check P$ is an indefinite Hermitian involution,
$\check{P}\check{T}$ is an antilinear involution (so that $[\check
P,\check T]=0$) and that $\check H$ has an exact
$\check{P}\check{T}$-symmetry. Because $\check T=\star$, the
matrices $\check P$ and $\check H$ have the general form
(\ref{e30}) and (\ref{H=3}) respectively. Therefore, according to
(\ref{e61}) the most general $2\times 2$ Hamiltonian admitting an
exact $PT$-symmetry such that $P$ and $T$ are general Hermitian
(respectively linear and antilinear) commuting involutions is
given by
    \be
    H=U~\check H\; U^{-1},
    \label{H5}
    \end{equation}
where $U$ is the unitary matrix (\ref{U=5}) and $\check H$ is
given by the right-hand side of (\ref{H=3}).

Because $\check H$ is a Hamiltonian of the form (\ref{H=3}),
according to (\ref{H=UH-tilde-U}) it may be mapped to a Hermitian
Hamiltonian of the form (\ref{tilde-H}) by a unitary
transformation. This observation together with Eq.~(\ref{H5}) and
the fact that $U$ is a unitary matrix, indicate that the most
general $2\times 2$ Hamiltonian having exact $PT$-symmetry is
related to a Hermitian $2\times 2$ Hamiltonian by a unitary
transformation.

In this article we showed that if the Hamiltonian of a quantum
system has an exact $PT$-symmetry and supports a unitary
time-evolution, then the same system may be described using a
Hermitian Hamiltonian. This provides the following answer to the
question: ``Must a Hamiltonian be Hermitian?'' posed in the title
of \cite{b3}: `The Hamiltonian need not be Hermitian in a given
inner product, but if one demands unitarity then one can describe
the same quantum system using a Hermitian Hamiltonian.'

If the Hilbert space is finite-dimensional, in general, there is
no practical difference between non-Hermitian Hamiltonians
supporting unitary evolutions and Hermitian Hamiltonians. For the
case that the Hilbert space is an infinite-dimensional function
space, again such a non-Hermitian Hamiltonian can be mapped to a
physically equivalent Hermitian Hamiltonian. But the latter is
generally a nonlocal (non-differential) operator. In other words,
a local Hamiltonian which is non-Hermitian with respect to a given
(positive-definite) inner product $\pbr~~,~~\pkt$ will support a
unitary evolution with respect to another (positive-definite)
$\br~~,~~\kt$ inner product if and only if it is physically
equivalent to a Hamiltonian which is Hermitian with respect to the
original inner product $\pbr~~,~~\pkt$. The only advantage of
exploring exact $PT$-symmetric (quasi-Hermitian) Hamiltonians is
that the corresponding equivalent Hermitian Hamiltonians may be
nonlocal operators whose study is generally more difficult. This
observation also suggests a similar scenario for the non-Hermitian
$CPT$-symmetric local field theories, namely that such a theory is
equivalent to a Hermitian nonlocal field theory. A direct
implication of this statement is that non-Hermitian
$CPT$-symmetric local field theories are expected to share the
appealing properties of nonlocal field theories, but since they
are local field theories they may prove to be much simpler to
study.

Finally, we wish to point out that one can also consider
time-dependent exact $PT$-symmetric (quasi-Hermitian)
Hamiltonians. The issue of the unitarity of the time-evolution for
this kind of Hamiltonians is more subtle. It plays an interesting
role in the solution of the Hilbert space problem for certain
quantum cosmological models \cite{cqg}.

\subsection*{Acknowledgment}

This work has been supported by the Turkish Academy of Sciences in
the framework of the Young Researcher Award Program
(EA-T$\ddot{\rm U}$BA-GEB$\dot{\rm I}$P/2001-1-1).


\ed
\begin{thebibliography}{99}
\bibitem{bender} C.~M.~Bender, S.~Boettcher,
Phys.\ Rev.\ Lett., {\bf 80}, 5243 (1998);\\
C.~M.~Bender, S.~Boettcher, and P.~N.~Meisenger, J.~Math.\ Phys.\
{\bf 40}, 2201 (1999).
\bibitem{p1} A. Mostafazadeh,
J.\ Math.\ Phys., {\bf 43}, 205 (2002); LANL ArXiv:
math-ph/0107001.
\bibitem{p2} A. Mostafazadeh,
J.\ Math.\ Phys., {\bf 43}, 2814 (2002); LANL ArXiv:
math-ph/0110016.
\bibitem{p3} A. Mostafazadeh,
J.\ Math.\ Phys., {\bf 43}, 3944 (2002); LANL ArXiv:
math-ph/0203005.
\bibitem{p4} A.~Mostafazadeh,
Nucl.\ Phys.\ B, {\bf 640}, 419 (2002); LANL ArXiv:
math-ph/0203041.
\bibitem{p5} A.~Mostafazadeh,
Mod.\ Phys.\ Lett.\ A, {\bf 17}, 1973 (2002); LANL ArXiv:
math-ph/0204013.
\bibitem{p6} A.~Mostafazadeh, J.\ Math.\ Phys., {\bf 43}, 6343
(2002); Erratum: {\bf 44}, 943 (2003); LANL ArXiv: math-ph/0207009
\& 0301030.
\bibitem{p7} A.~Mostafazadeh,
J.\ Math.\ Phys., {\bf 44}, 974 (2003); LANL ArXiv:
math-ph/0209018.
\bibitem{p8} A.~Mostafazadeh, `Pseudo-Unitary Operators and
Pseudo-Unitary Quantum Dynamics,' LANL ArXiv: math-ph/0302050.
\bibitem{others}
Z.\ Ahmed, Phys.\ Lett.\ A, {\bf 290}, 19 (2001); ibid {\bf 294},
287 (2002);\\
B.\ Bagchi and C.\ Quesne, Phys.\ Lett.\ A, {\bf 301}, 173
(2002);\\
G.\ Scolarici and L.\ Solombrino, Phys.\ Lett.\ A, {\bf 303},
239-242 (2002);\\
G.\ Scolarici,  J.\ Phys.\ A: Math.\ Gen., {\bf 35}, 7493
(2002);\\
L.\ Solombrino, J.~Math.\ Phys., {\bf 43}, 5439 (2002);\\
M.~Znojil, `Pseudo-Hermitian version of the charged harmonic
oscillator and its ``forgotten" exact solutions,' LANL ArXiv:
quant-ph/0206085;\\
G.\ Scolarici and L.\ Solombrino, `On the pseudo-Hermitian
nondiagonalizable Hamiltonians,' LANL ArXiv: quant-ph/0211161.
\bibitem{bmw1} C.~M.~Bender, P.~N.~Meisenger, and Q.~Wang,
J.\ Phys.\ A: Math.\ Gen., {\bf 36}, 1029 (2003); LANL ArXiv:
quant-ph/0211123.
\bibitem{bmw2} C.~M.~Bender, P.~N.~Meisenger, and Q.~Wang,
J.\ Phys.\ A: Math.\ Gen., {\bf 36}, 6791 (2003); LANL ArXiv:
quant-ph/0303174.
\bibitem{kato} T.\ Kato, {\em Perturbation Theory for Linear
Operators} (Springer, Berlin, 1995).
\bibitem{nova} A.~Mostafazadeh, {\em Dynamical Invariants,
Adiabatic Approximation, and the Geometric Phase} (Nova Science
Publishers, New York, 2001).
\bibitem{quasi} F.\ G.\ Scholtz, H.\ B.\ Geyer, and
F.\ J.\ W.\ Hahne, Ann.\ Phys.\ N.~Y.\ {\bf 213}, 74 (1992).
\bibitem{weinberg} S.~Weinberg, {\em The Quantum Theory of Fields}, Vol.~I (Cambridge University Press, Cambridge, 1995).
\bibitem{b3} C.~M.~Bender, D.~C.~Brody, and H.~F.~Jones, `Must a
Hamiltonian Be Hermitian?' LANL ArXiv: hep-th/0303005.
\bibitem{cqg} A.~Mostafazadeh,
Class.\ Quantum Grav.\ {\bf 20}, 155 (2003); LANL ArXiv:
math-ph/0209014.
\end{thebibliography}
